# Common Frame of reference in collaborative virtual environments and their impact on presence

Amine Chellali[1, 2], Cedric Dumas[1, 2], Isabelle Milleville-Pennel[2], Eric Nouri[3]
[1]EMN, [2]IRCCyN CNRS-UMR 6597, [3]Université de Nantes
{amine.chellali, cedric.dumas}@emn.fr, isabelle.Milleville-Pennel@irccyn.ec-nantes.fr, nourieric@hotmail.com

**Abstract**

*Virtual collaborative environment are 3D shared spaces in which people can work together. To collaborate through these systems, users must have a shared comprehension of the environment. The objective of this experimental study was to determine if visual stable landmarks improve the construction of a common representation of the virtual environment and thus facilitate collaboration. This seems to increase the awareness of the partner's presence.*

## 1. Introduction

Collaborative virtual environments (CVE) are digital spaces in which distant users can meet, share virtual objects and work together. CVE can be used in many areas like e-learning, training and entertainment [1]. We are particularly interested in the use of the CVE for technical gesture learning and virtual objects co-manipulation.

In a collaborative environment, when users are physically distant, the channels of communication are restricted. This may penalize human-human collaboration, especially when users have to manipulate together virtual objects. Thus, users have to get a common representation of the virtual world, so that they avoid mutual incomprehension.

In this paper we will define the notion of common Frame of Reference (FR), and will present an experimental study that shows how operators can build common representations to collaborate when they do not share the same view of a VE.

## 2. Common frame of reference

When persons have to perform a shared task, they need both a shared representation of their actions on these objects and a common spatial representation [2]. These representations permit compatible decisions. User's action on an object can be specified using two different spatial frames of reference for action [3]:
(a) An egocentric FR, in which locations are represented with respect to the perspective of a user;
(b) An allocentric FR that locates points within a framework external to the user (stable visual landmarks SVL) and independent of his or her position (viewpoint).

In 3D VE, users tend to use an egocentric FR to plan their actions. For example, to recognise a 3D object shape and plan to manipulate it, an operator makes a mental rotation to fit the object viewpoint to his own viewpoint [4]. This allows him to construct his personal comprehension of the environment. However, in a collaborative task, egocentric reference frame (which is specific to each operator) may worsen the share of the personal comprehension and then restrain the elaboration a common spatial representation of the VE. Thus, the use of an allocentric FR enlarges the common spatial FR (since objects are located independently from operator viewpoints).

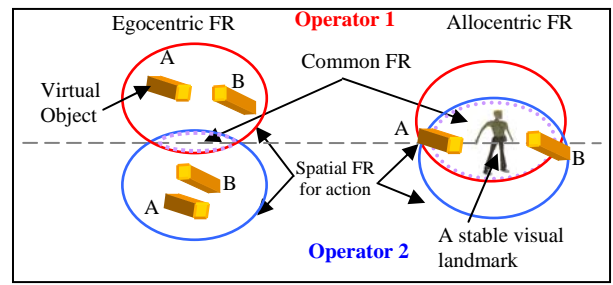

**Figure 1: Egocentric (left) and Allocentric (right) FR in collaborative environment**

Some research studies tried to increase collaboration by allowing the sharing of the collaborator's viewpoint [5]. However these studies did not try to improve mutual comprehension of actions and intentions.

In this experimental study, we focused on the effects of the SVL presence in a shared VE and how it can be used to construct a common FR between peers. To isolate these effects, no other indications (viewpoint and/or position of the other user) were given. We also study how the use of a virtual character as a SVL can affect the awareness of the collaborator's presence in the VE.

## 3. Method

### 3.1. Hypotheses

In presence of a SVL, the operators use an allocentric FR, thus they can get a larger common spatial FR and collaboration becomes easier: The collaboration is evaluated according to mutual understanding between peers and not according to the task's completion time.

In presence of the SVL, operators presence awareness of each other increases since they tend to work together.

Men and women have different spatial abilities especially to perform mental rotations as suggested in [6].

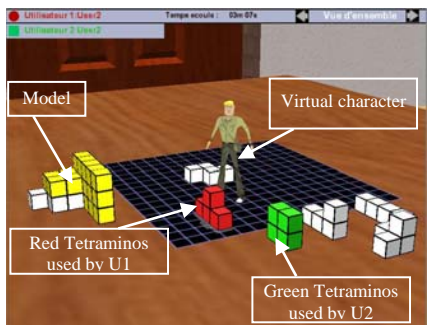

**Figure 2: the virtual interface**

### 3.2 Interface and Procedure

Forty four students (20-27 years old) participated in the experiment. The subjects performed the experiment in 22 pairs (10 female pairs and 12 male pairs).

The collaborative VE consists of a table on which laid a model to reproduce and 6 different white tetraminos[1] (see figure 2). Tetraminos can be moved using a gamepad. Two users can move two different tetraminos at the same time.

Each subject was seated in front of a computer screen and held a gamepad. The two participants were in the same room but could not see each other's screen. However, they were encouraged to communicate verbally.

The subjects were asked to reproduce together models using the 6 tetraminos. The starting viewpoints of subjects were different from each other. However, each participant was allowed to change his viewpoint during the task (turning around the scene). Two experimental conditions were tested:
(i) A 3D character (SVL) was placed in the scene's center.
(ii) There was no virtual 3D character (egocentric system).

### 4. Results:

The results showed no significant differences between character presence condition and character absence condition. However the difference in completion time was observed between male pairs and female pairs: female pairs took much more time (620 sc) than male pairs (380 sc).

The results showed that users spent much more time together in the same viewpoint in character's presence condition. However, this time represents a small percentage of the total time (less than 40%).

The ongoing verbalisations studies (percentage of pronouns correctly resolved, location of objects and actions)

---

[1] A tetramino is a geometric figure composed of 4 cubes

indicate that operators spontaneously used an allocentric FR in the character presence condition.

### 5. Discussion and conclusion

The results indicate that it takes more time to female subjects to perform the task than male subjects; however there were no significant differences between character presence and character absence conditions. These results are consistent with those of Kimura [6] and suggest a difference in performing mental rotations between men and women. No completion time difference was expected between character presence and character absence conditions, since the task can be accomplished individually as well as in collaboration.

First verbalisations analyses show that operators tend to use an allocentric FR when the virtual character is present. Thus, operators spent little time together in the same viewpoints, since the construction of the Common FR is independent from viewpoint.

In the character absence condition, the operators spent also little time together in same viewpoints. In fact, in absence of a SVL, operators used an egocentric FR and so the definition of common FR became harder. These findings are confirmed by verbalisations analyses, since the task resolutions were more individual in this condition.

In this study no representation of operators in VE was used and only one indicator of presence was used (the colour of the tetraminos being manipulated). However, we observed that operators acted like if the tetraminos manipulated were the representations of the manipulator in the VE. This suggests that a simple user representation in the VE can be sufficient to be aware of the presence of other users.

More investigations are being made on verbalisations to determine whether or not the use of a SVL has an indirect impact on the awareness of the partner's presence.